# SHM method for damage localization based on substructuring and VARX models

U. Ugalde[1], J. Anduaga[1], F. Martínez[1] and A. Iturrospe[2]

[1] *Department of Sensors, IK4-Ikerlan, Mondragon, Spain (uugalde@ikerlan.es)*
[2] *Department of Electronics and Computer Sciences, Mondragon Goi Eskola Politeknikoa, Mondragon, Spain*

*Abstract—* A novel damage localization method is proposed, which is based on a substructuring approach and makes use of Vector Auto-Regressive with eXogenous input (VARX) models. The substructuring approach aims to divide the monitored structure into several multi-DOF isolated substructures. Later, each individual substructure is modeled by a VARX model, and the health of each substructure is determined analyzing the variation of the VARX model. The method allows to detect whether the isolated substructure is damaged, and besides allows to locate the damage within the substructure. Only measured displacement data is required to estimate the isolated substructure's VARX model. Moreover, it is not necessary to have *a priori* knowledge of the structural model. The proposed method is validated by simulations of an eight-storey shear building.

## 1 INTRODUCTION

Structural Health Monitoring (SHM) is the process of implementing a damage detection and characterization strategy for engineering structures [1]. SHM is regarded as a very important engineering field in order to secure structural and operational safety; issuing early warnings on damage or deterioration, avoiding costly repairs or even catastrophic collapses [2].

Most of the existing vibration based SHM methods could be classified into two different approaches: global approaches and local approaches [3]. In the global approaches, the goal is to monitor the health of the entire structure. These global methods have been tested and implemented in different types of structures during the last 30 years [4]. However, for many large systems, global monitoring is not practical due to the lack of sensitivity of global features regarding local damages, inaccuracies of developed models or the high cost of sensing, cabling and computational operations [5]. On the other hand, local SHM methods are focused on evaluating the state of reduced parts within the entire structures, based on substructuring methods. This approach aims to overcome global method's problems, dividing the whole structure into substructures and analyzing each one individually.

Several research works have proposed substructuring methods for large-scale structures. Koh [6] presented a "divide and conquer" strategy to monitor large structures based on the division of the whole structure into isolated substructures. For each substructure, structural parameters are identified using the Extended Kalman filter (EKF). However, the EKF usually require a previous knowledge of the system and its dynamics [7]. Yun and Lee [8] detected damage in structures combining a substructuring method and Auto-Regressive Moving Average with eXogenous input (ARMAX) models. Most recently, Xing [9] presented another damage detection method based on a substructuring method and ARMAX models. Damage indicators were obtained for each estimated substructure model, by calculating the difference between the squared natural frequencies in the healthy state and the squared natural frequencies during the structure lifetime. All the natural frequencies were computed from their respective estimated ARMAX models. This damage detection method was validated through simulations and experimental test. The method proposed by Xing [9] doesn't require any previous knowledge of the structure. Nevertheless, as only one internal DOF is measured in each substructure, is not possible to give information about the damage location within the substructures.

In this paper, a damage localization method based on the combination of a substructuring method and Vector Auto-Regressive with eXogenous input (VARX) models is proposed. The substructuring method is used to isolate a multi-DOF substructure from the rest of the structure, and each isolated substructure is modeled by a VARX model. VARX models incorporate data measured in different internal DOFs and their coefficient matrices describe the relationship between the measured internal DOFs through some structural features (mass, stiffness, damping…). Therefore, the proposed method can potentially locate the damage within the substructure by analyzing variations on the VARX model over the time. Furthermore, the proposed method doesn't require a precise *a priori* knowledge of the structure.

The rest of the paper is organized as follows. First, the proposed method is presented in section 2. Secondly, the proposed method is evaluated by series of simulations. In section 3, simulation results are discussed and finally the concluding remarks are presented in section 4.

## 2 THE PROPOSED METHOD

A multi-DOF shear structure is modeled as a one-dimensional lumped mass-spring model (Figure 1). Mass and stiffness values are respectively $m_j$ and $k_j$, and $z_j$ represents the absolute displacement associated to each DOF. The structure is divided into $m$ different substructures, where each substructure contains several internal DOFs and two interface DOFs.

### 2.1. Substructure's dynamic equation

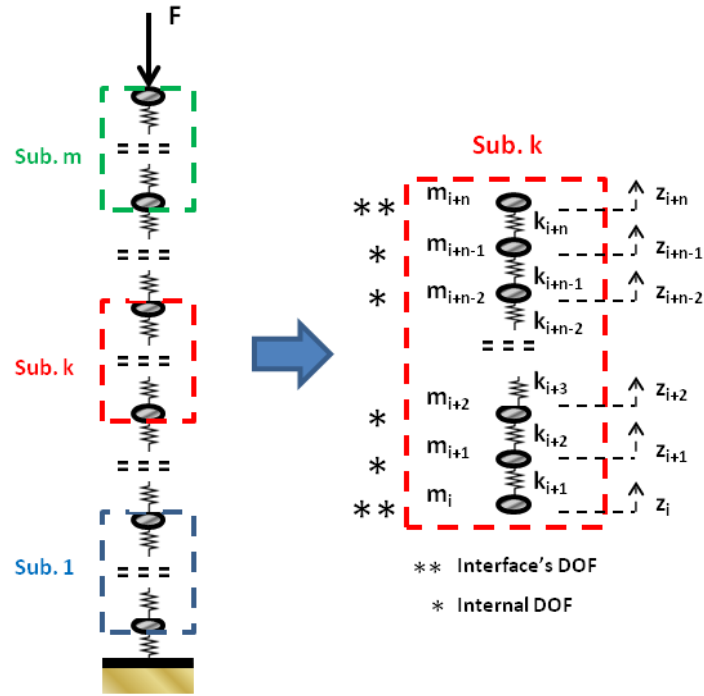

**Figure 1: Simplified structural model**

The dynamic equation for the substructure $k$, which consists of $n$ DOFs, could be formulated as:

$$m_{i+1}\ddot{z}_{i+1}(t) = -k_{i+1}(z_{i+1}(t) - z_i(t)) - k_{i+2}(z_{i+1}(t) - z_{i+2}(t)) \quad (1)$$

$$\ldots$$

$$m_{i+n-1}\ddot{z}_{i+n-1}(t) = -k_{i+n-1}(z_{i+n-1}(t) - z_{i+n-2}(t)) - k_{i+n}(z_{i+n-1}(t) - z_{i+n}(t))$$

where $\ddot{z}_j$ are absolute accelerations and $z_j$ are the absolute displacement associated to each DOFs respectively.

Discretizing by the central difference approximation the Equation (1), the substructure dynamics are stated as:

$$\frac{m_{i+1}}{T_s^2}z_{i+1}(n+1) - 2\frac{m_{i+1}}{T_s^2}z_{i+1}(n) + \frac{m_{i+1}}{T_s^2}z_{i+1}(n-1) = -k_{i+1}(z_{i+1}(n) - z_i(n)) - k_{i+2}(z_{i+1}(n) - z_{i+2}(n)) \quad (2)$$

...

$$\frac{m_{i+n-1}}{T_s^2}z_{i+n-1}(n+1) - 2\frac{m_{i+n-1}}{T_s^2}z_{i+n-1}(n) + \frac{m_{i+n-1}}{T_s^2}z_{i+n-1}(n-1) = -k_{i+n-1}(z_{i+n-1}(n) - z_{i+n-2}(n)) - k_{i+n}(z_{i+n-1}(n) - z_{i+n}(n))$$

Finally, representing the discrete dynamic equations in matrix form, the next expression is obtained:

$$\begin{bmatrix} z_{i+1}(n) \\ z_{i+2}(n) \\ ... \\ z_{i+n-2}(n) \\ z_{i+n-1}(n) \end{bmatrix} = -\begin{bmatrix} \frac{T_s^2}{m_{i+1}}(-\frac{2m_{i+1}}{T_s^2} + k_{i+1} + k_{i+2}) & -T_s^2\frac{k_{i+2}}{m_{i+1}} & ... & 0 & 0 \\ ... & ... & ... & ... & ... \\ ... & ... & ... & ... & ... \\ ... & ... & ... & ... & ... \\ 0 & 0 & ... & -T_s^2\frac{k_{i+n-1}}{m_{i+n-1}} & \frac{T_s^2}{m_{i+n-1}}(-\frac{2m_{i+n-1}}{T_s^2} + k_{i+n-1} + k_{i+n}) \end{bmatrix}\begin{bmatrix} z_{i+1}(n-1) \\ z_{i+2}(n-1) \\ ... \\ z_{i+n-2}(n-1) \\ z_{i+n-1}(n-1) \end{bmatrix} - \quad (3)$$

$$-\begin{bmatrix} 1 & 0 & ... & 0 & 0 \\ 0 & 1 & ... & 0 & 0 \\ 0 & 0 & ... & 0 & 0 \\ 0 & 0 & ... & 1 & 0 \\ 0 & 0 & ... & 0 & 1 \end{bmatrix}\begin{bmatrix} z_{i+1}(n-2) \\ z_{i+2}(n-2) \\ ... \\ z_{i+n-2}(n-2) \\ z_{i+n-1}(n-2) \end{bmatrix} +$$

$$+\begin{bmatrix} T_s^2\frac{k_{i+1}}{m_{i+1}} & 0 \\ ... & ... \\ ... & ... \\ ... & ... \\ 0 & T_s^2\frac{k_{i+n}}{m_{i+n-1}} \end{bmatrix}\begin{bmatrix} z_i(n-1) \\ z_{i+n}(n-1) \end{bmatrix}$$

Equation (3) could be regarded as a two exogenous and (*n-2*) endogenous variables VARX model [10]. Equation (3) shows that the elements of all coefficient matrices are function of the sampling period, as well as the mass and stiffness of the structural elements.

## 3 SIMULATIONS

A linear and time invariant eight-story shear building is modeled as an eight DOF lumped mass-spring model, as shown in the Figure 2. The mass of every floor is 100 kg and the stiffness of every spring is 1M N/m.

In the studied case, a five DOF substructure is isolated from the general structure as explained in section 2. As shown in Figure 2, displacement $z_3$, $z_4$, and $z_5$ correspond to internal DOFs and displacement responses $z_2$ and $z_6$ correspond to interface DOFs.

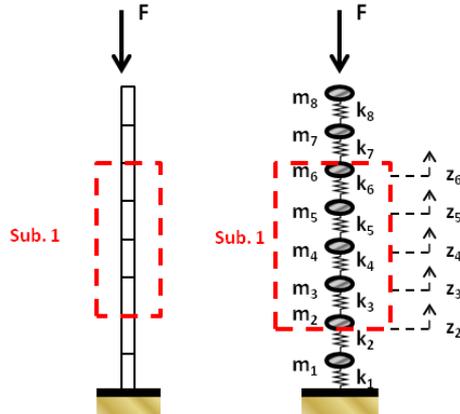

**Figure 2: Isolated substructure in an eight story shear building model**

Below, the dynamic equation for the isolated substructure is formulated:

$$m_3\ddot{z}_3(t) = -k_3(z_3(t) - z_2(t)) - k_4(z_3(t) - z_4(t))$$ (4)
$$m_4\ddot{z}_4(t) = -k_4(z_4(t) - z_3(t)) - k_5(z_4(t) - z_5(t))$$
$$m_5\ddot{z}_5(t) = -k_5(z_5(t) - z_4(t)) - k_6(z_5(t) - z_6(t))$$

Following the procedure described in section 2, we get the VARX model. Equation (5) could be regarded as a two exogenous and three endogenous variables VARX model [10]. The exogenous variables are the measured displacements in DOFs 2 ($z_2$) and 6 ($z_6$), and the endogenous variables are the measured displacement in DOFs from 3 ($z_3$) to 5 ($z_5$). $A_1$ and $A_2$ are endogenous coefficient matrices and $B_1$ is an exogenous coefficient matrix. Equation (5) shows that $A_1$ is a 3 x 3 matrix and its elements are function of substructural parameters like masses and springs ($m_3$, $m_4$, $m_5$, $k_3$, $k_4$, $k_5$, $k_6$).

$$\begin{bmatrix} z_3(n) \\ z_4(n) \\ z_5(n) \end{bmatrix} = - \begin{bmatrix} \frac{T_s^2}{m_3}(-\frac{2m_3}{T_s^2} + k_3 + k_4) & -T_s^2 \frac{k_4}{m_3} & 0 \\ -T_s^2 \frac{k_4}{m_4} & \frac{T_s^2}{m_4}(-2\frac{m_4}{T_s^2} + k_4 + k_5) & -T_s^2 \frac{k_5}{m_4} \\ 0 & -T_s^2 \frac{k_5}{m_5} & \frac{T_s^2}{m_5}(-\frac{2m_5}{T_s^2} + k_5 + k_6) \end{bmatrix} \begin{bmatrix} z_3(n-1) \\ z_4(n-1) \\ z_5(n-1) \end{bmatrix} - $$ (5)

$$- \begin{bmatrix} 1 & 0 & 0 \\ 0 & 1 & 0 \\ 0 & 0 & 1 \end{bmatrix} \begin{bmatrix} z_3(n-2) \\ z_4(n-2) \\ z_5(n-2) \end{bmatrix} +$$

$$+ \begin{bmatrix} T_s^2 \frac{k_3}{m_3} & 0 \\ 0 & 0 \\ 0 & T_s^2 \frac{k_6}{m_5} \end{bmatrix} \begin{bmatrix} z_2(n-1) \\ z_6(n-1) \end{bmatrix}$$

The structure is excited in the top outside the substructure by a Gaussian white noise force and the displacements are recorded for each substructure DOF using a data sampling frequency of 1000 Hz. The substructure VARX model is estimated by the Multivariable Least-Square estimator (MLS) method [11] both for a healthy state and several damaged scenarios described in Table 1. All considered damages are stiffness losses of one specific spring within the structure model. Three different damage severities (5%, 10% and 20%) and six different damage locations are assessed. In some of them, the damaged springs are within the substructure ($k_3$, $k_4$, $k_5$, $k_6$) and in the others, they correspond to external spring ($k_1$, $k_8$).

As we can see in Equation (5), some of the elements of the matrix $A_1$ ($A_{1(1,1)}$, $A_{1(1,2)}$, $A_{1(2,1)}$, $A_{1(2,2)}$, $A_{1(2,3)}$, $A_{1(3,2)}$, and $A_{1(3,3)}$) are function of the substructural springs's stiffness values ($k_3$, $k_4$, $k_5$, $k_6$). In this work, we calculate a damage indicator value (DI) for these elements in every new scenario. The DI values are calculated depending on the variation that the elements ($A_{1(1,1)}$, $A_{1(1,2)}$, $A_{1(2,1)}$, $A_{1(2,2)}$, $A_{1(2,3)}$, $A_{1(3,2)}$, and $A_{1(3,3)}$) have had respect to their healthy values. In order to locate damages within the substructure, these DI values are analyzed. Table 1 shows which $A_1$'s elements should change in each studied scenario.

**Table 1: Assessed damaged scenarios**

| Scenario number | Modified spring | Within substructure | Hypothetically affected $A_1$'s elements |
|---|---|---|---|
| 1 | $k_1$ | No | - |
| 2 | $k_3$ | Yes | $A_{1(1,1)}$ |
| 3 | $k_4$ | Yes | $A_{1(1,1)}$, $A_{1(1,2)}$, $A_{1(2,1)}$, $A_{1(2,2)}$ |
| 4 | $k_5$ | Yes | $A_{1(2,2)}$, $A_{1(2,3)}$, $A_{1(3,2)}$, $A_{1(3,3)}$ |
| 5 | $k_6$ | Yes | $A_{1(3,3)}$ |
| 6 | $k_8$ | No | - |

Regarding to the results, for external damages (reducing $k_1$ and $k_8$ values), the calculated DI values for all elements ($A_{1(1,1)}$, $A_{1(1,2)}$, $A_{1(2,1)}$, $A_{1(2,2)}$, $A_{1(2,3)}$, $A_{1(3,2)}$, and $A_{1(3,3)}$) are almost zero, so the method determines that the substructure is healthy. For internal damages (reducing $k_3$, $k_4$, $k_5$ and $k_6$ values), the DI values are shown in Figure 3. Furthermore, DI values give information about the damaged spring within the substructure, as well as the damage severity.

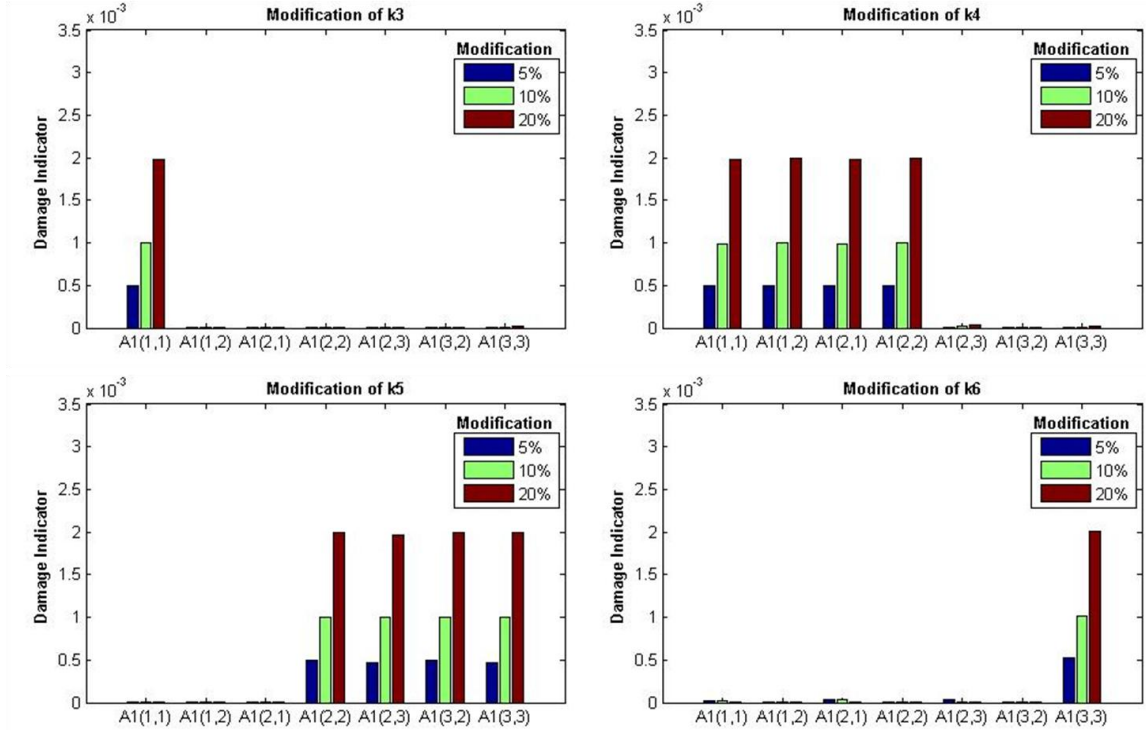

Figure 3: DI values for the analyzed $A_1$'s elements (internal damages)

## 4 CONCLUSIONS

This paper proposes a novel SHM method to locate damages in structures. A substructure of interest is isolated by a substructuring method and a VARX model of the isolated substructure is obtained. The analysis of the estimated VARX model is carried out in order to assess the health of the isolated substructure. Only measured displacement data is required to estimate the VARX model and it is not necessary to have *a priori* knowledge of the structure.

A linear and time invariant model for an eight story shear building is simulated to evaluate the proposed method. The results show that the method not only allows detecting damage within the substructure, it also estimates both the damage severity and the damage location within the substructure.

The proposed method is also suited for 2D and 3D lattice structures, where the number of element's connections increases. Our research group is already applying this method in 2D and 3D structures and the results will be published soon.